 \documentstyle[prb,aps,epsfig,multicol]{revtex} 
 \renewcommand{\narrowtext}{\begin{multicols}{2} \global\columnwidth20.5pc}
 \renewcommand{\widetext}{\end{multicols} \global\columnwidth42.5pc}

\begin{document}

\draft 

\title{Two-dimensional Non-Hermitian Delocalization Transition as a
  Probe for the Localization Length}
 
\author{Tsunenao Kuwae$^{a}$ and Nobuhiko Taniguchi$^{b}$}

\address{$^{a}$Department of Physical Electronics, Hiroshima
  University,  Kagamiyama, Higashi-Hiroshima 739-8527, Japan \\ %
  $^{b}$Institute of Physics, University of Tsukuba, Tennodai Tsukuba
  305-8571, Japan}

\date{\today}

\maketitle%

\begin{abstract}
  When one applies a type of non-Hermitian effect, constant imaginary
  vector potential, to disordered systems, delocalization is induced
  even in two or lower dimension.  By using the non-Hermitian induced
  transition as a probe, We propose a new procedure of estimating
  localization in arbitrary-dimensional systems.  By examining
  numerically the two-dimensional non-Hermitian tight-biding model
  with onsite disorder, it is shown that the failure of absorbing the
  non-Hermitian effect, namely the breakdown of the imaginary gauge
  transformation, characterize the inverse localization
  length near the band center.
\end{abstract}%


\narrowtext


In non-interacting disordered systems with time-reversal symmetry, all
the states are believed to be localized in two or lower dimensions.
Often one encounters situations where non-Hermitian effect
emerges effectively through coupling with the reservoir, interaction
between particles, or non-equilibrium relaxation processes.  This type
of non-Hermitian effect is taken into account as a finite inelastic
scattering time, hence following the scaling theory idea, localization
will be truncated halfway only to exhibit a gradual {\em crossover}
into metallic behavior.
Recently, in a vortex pinning problem in a bulk superconductor by
columnar defects, it has been stressed that the non-Hermitian term
coupled with the first-order differential, {\it i.e.}, the imaginary
vector potential, gives a fundamentally different effect on
localization states in disordered systems~\cite{Hatano9697}.  The term
appears to take account for the in-plane magnetic field component.
Along with subsequent numerical and analytical
studies~\cite{Brouwer97,Goldsheid98,Silvestrov98}, it has been
recognized that sufficiently large constant imaginary vector potential
induces states with complex eigen energies even in one dimension.  The
phenominon has later been understood as a non-Hermitian induced
delocalization {\em transition} not a crossover.  The same class of
non-Hermitian terms is ubiquitous in diverse problems: classical
(elastic, electromagnetic, or acoustic) wave propagation with
absorptive medium\cite{Freilikher94}, random walks in random media
with drifting forces~\cite{randomFP}, or Burgers turbulence through a
Cole-Hopf transformation~\cite{Bouchaud95}.

While non-Hermitian effect is usually hard to control, the strength
of the imaginary vector potential is relatively easy to regulate not
only in numerical studies but also in experiments.  For instance, one
changes the tilt angle of the magnetic field in the vortex pinning
problem, or tunes an incident frequency and change the absorption rate
in electromagnetic wave propagation.  It has been suggested that the
non-Hermitian induced delocalization process can be detected by the
energy spectrum.  So with its drastic consequence on localization, the
phenomenon provides a great opportunity to investigate the
localization property without resorting to any transport phenomenon,
{\it i.e.}, neither attaching leads nor coupling with the reservoir.

In this paper, pursuing the above possibility, we propose a new
procedure (Eq.~(\ref{eq:newrel}) below) appropriate to extract the
inverse localization length out of the non-Hermitian dependence of the
energy spectrum.  Whereas the connection between the average density
of states (DOS) and the localization length has been successfully
managed in one dimension, it is not the case in higher
dimensions~\cite{Mudry98b} (See the argument below).  Our proposed
procedure is principally applicable in arbitrary-dimensional systems.
By using the two-dimensional (2D) non-Hermitian model with
time-reversal symmetry, we will confirm our proposal numerically.  The
non-Hermitian delocalization in 2D systems is far from complete
understanding, and it is still at issue whether the transition is
continuous or abrupt~\cite{Mudry98,Kolesnikov00}.  On this regard, we
also briefly mention our point of view based on numerical results.

The Hamiltonian considered for numerical calculations is the
$d$-dimensional non-Hermitian tight-binding model,
\begin{equation}
  H(\bbox{h}) = -{t \over 2}\sum_{(i,j)} \exp \left( {\bbox{h}\cdot
      \bbox{r}_{ij}} \right)\,|i\rangle \langle j| + \sum_{i} V_{i}\,
      |i\rangle \langle i|,
 \label{def:Hamiltonian}
\end{equation}
where $\bbox{h}$ is the directed hopping, {\it i.e.}, a lattice analog
of the imaginary vector potential.  $\{V_{i}\}$ is a random on-site
potential chosen from a box distribution over $(-W,W)$.  $|i\rangle$
denotes the state bound at $\bbox{r}_{i}$, and the lattice summation
$(i,j)$ is restricted over nearest-neighboring pairs of sites.  We
work on the two-dimensional case, where we particularly choose
$\bbox{h}$ as $h_{x}=h_{y} = h$ to avoid unnecessary troubles from
lattice periodicity.  In the rest of the paper, we adopt
the convention of the unit lattice constant and $t=2$ as well as
$\hbar=1$.
Using an exact diagonalization technique, we have calculated a
complete set of eigen values and eigen wavefunctions on a
two-dimensional $20\times 20$ lattice.  (Eigen functions are needed
later to justify our interpretation.)  Ensemble average is taken over
40 samples.

The possibility to detect localization by using the non-Hermitian
dependence of the spectrum has already been noticed in
Ref.~\cite{Hatano9697}.  The reasoning is as follows.  Suppose an
eigen state $\phi_{\bbox{h}=\bbox{0}}(\bbox{r})$ be localized
exponentially around $\bbox{r}_{0}$ with the localization length
$\xi$, that is, $\phi_{\bbox{0}} \sim
e^{-|\bbox{r}-\bbox{r}_{0}|/\xi}$.  When one turns on $\bbox{h}$
gradually, its eigen energy does not change in value, while the right
eigen wave function progresses following the {\em imaginary} gauge
transformation $\phi_{\bbox{h}} (\bbox{r}) \sim e^{\bbox{h} \cdot
  (\bbox{r}-\bbox{r}_{0})} \phi_{\bbox{0}} (\bbox{r})$ up to a small
correction by the boundary condition.  Such `absorption' of the
non-Hermitian influence on the spectrum can take effect only when
$\bbox{h}$ is smaller than the inverse localization length.  Beyond
that, the state undergoes such a drastic change that it transforms
into a complex energy (delocalized) state.  Putting the other way
round, one uses the breakdown of the imaginary gauge transformation to
characterize and give a practical estimate of the inverse
localization length.

A naive way to estimate the breakdown of the imaginary gauge
transformation is to examine the average DOS defined on the complex
energy plane, $\rho_{h}(E,{\cal E})= \langle \delta(E-{\rm Re}E_{n})
\delta({\cal E}-{\rm Im}E_{n})\rangle$.  When all the states are
localized at sufficiently small $h$, energies are all real and
$\rho_{h}(E,{\cal E}) \propto \delta({\cal E})$.  With increasing $h$,
the support of $\rho_{h}(E, {\cal E})$ extends into a ${\cal E}\neq 0$
region in the presence of complex energy states.  By defining
$h_{0}(E)$ by the point having a finite density at ${\cal E}\neq 0$
for each $E$ one sees $h_{0}(E)$ indicate the breakdown of the
imaginary gauge transformation.  Accordingly, it is suggested that the
inverse localization length may be estimated by the formula $1/\xi(E)=
h_{0}(E)$~\cite{Hatano9697,GurarieBrezin98}.

While the above argument is conceptually simple and attractive, the
estimate of the localization length by using the average DOS has been
successful only in one dimension, whose success is traced to a special
structure of the energy spectrum.  In one dimension, irrespective of
the value of $h$, the support of $\rho_{h}(E,{\cal E})$ becomes
one-dimensional on the complex energy plane, {\it i.e.}, consisting of
two line segments along the real axis plus a `bubble' part in-between
corresponding to complex energy
states~\cite{Hatano9697,Brouwer97,Goldsheid98,Silvestrov98}.  The
shape of the bubble turns out quite insensitive to disorder
configurations.  Because of it, one can readily read off the value of
$h_{0}(E)$ from the endpoints of the arcs of the bubble.

\begin{figure}
  \centerline{\psfig{file=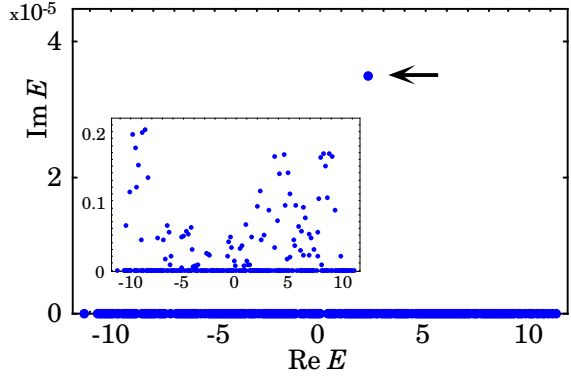,width=0.9\columnwidth}}
\caption
{An typical example of the energy spectrum for a given configuration
  of random site energies (only the upper half plane shown here).
  Random on-site potential is chosen from a box distribution
  $[-10,10]$.  The value of $h$ is adjusted to $h=0.21$, where only
  one pair of states (indicated by the arrow) get complex.  Inset: A
  typical structure of the DOS for a larger $h$ ($h=0.60$).  Both
  localized and delocalized states are mixed up in the same energy
  range. }
\label{fig:DOS}
\end{figure}

By contrast, the DOS support on higher dimensions is no longer
one-dimensional but much more involved.  A typical structure of the
two-dimensional DOS for a sufficiently large $h$ is shown in the inset
of Fig.~\ref{fig:DOS}.  As one sees, localized and delocalized states
coexist unpredictably over the entire energy range, showing that DOS
suffers large fluctuations.  The system has no clear-cut mobility edge
distinguishing between real-energy (localized) and complex-energy
(delocalized) states.  In fact, judging from existing
numerical~\cite{Hatano9697} and analytical~\cite{Efetov97} results, it
is very likely that the average DOS $\rho(E,{\cal E})$ {\em always}
retains a finite contribution proportional to $\delta({\cal E})$ hence
real energy states persist in existing for sufficiently large $h$,
even near the band center.
The feature of $\rho_{h}(E,{\cal E})$ in higher dimensions makes it
highly dubious to seek the breakdown of the imaginary gauge
transformation at a fixed $E$.  Moreover, numerical calculations on
two-dimensional systems suggest that $h_{0}(E)$ will estimate an
inverse localization length at more than one order of magnitude larger
than that obtained by any other method such as the recursion
method~\cite{MacKinnon83}.

Resolving problems above to answer whether the non-Hermitian effect
characterizes the localization length or not, we propose an
alternative criterion to extract the localization length.  Our
solution is to find the breakdown of the imaginary gauge
transformation in a more faithful and careful way: First we identify
the smallest directed hopping strength $h_{c}$ where {\em any} of the
eigen energies gets complex in a certain configuration of the random
potential (see Fig.~\ref{fig:DOS}), and next we take the ensemble
average of $h_{c}$ to obtain the inverse localization length
by~\cite{note1}
\begin{equation}
  \left\langle h_{c} \right\rangle = 1/\xi(0).
  \label{eq:newrel} 
\end{equation}
The estimated localization length corresponds to that around the band
center, since states near the band center tend to be delocalized
first.  Eq.~(\ref{eq:newrel}) and the procedure leading to it is our
main result in the paper.  The advantage of Eq.~(\ref{eq:newrel}) is
clear.  In any localized system, we can assign an unambiguous value of
$h_{c}$ for each configuration, and perform the ensemble average of
$h_{c}$.  It should be stressed that although there is no discrepancy
between $\langle h_{c}\rangle$ and $h_{0}(0)$ for one dimension,
$\langle h_{c}\rangle$ becomes much smaller than $h_{0}(0)$ in higher
dimensions, since the latter cannot detect a weakly-broken situation
as in Fig.~\ref{fig:DOS}.  Unlike $h_{0}(E)$ (if at all correctly
defined in higher dimensions), we cannot obtain $\langle h_{c}\rangle$
from the average DOS.

\begin{figure}
  \centerline{\psfig{file=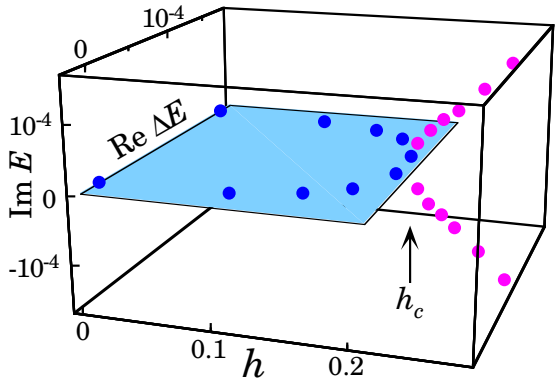,width=0.9\columnwidth}}
\caption
{ An evolution of the first pair of levels turning complex values, by
  increasing $h$.  In this configuration, we assign the critical value
  $h_{c}$ as $h_{c} = 0.21$. The ${\rm Im}E=0$ plane is drawn as a
  shaded plane for the eye guide.}
\label{fig:shiftedE}
\end{figure}

It is worth discussing a little more the definition of $h_{c}$ and
finite volume effect.  Although we have initially intended to assign
$h_{c}$ as the point where the imaginary gauge transformation fails to
absorb all the non-Hermitian effect, we should keep in mind that such
mechanism works only approximately in a finite system.  Figure
\ref{fig:shiftedE} represents a typical evolution of the first pair of
levels getting complex.  It shows that complex energy states appear
through merging two adjacent levels, but well before it, the pair of
levels shift gradually without gaining any imaginary part.  Strictly
speaking, the effect of $h$ remains partially for any small $h$, hence
the gauge transformation is already broken down for infinitesimal $h$.
Nevertheless, we assign the value of $h_{c}$ by a merging point.  The
shift of energies before $h_{c}$ arises from finite size effect, the
consistency with the periodic boundary condition in a finite system.

\begin{figure}
  \centerline{\psfig{file=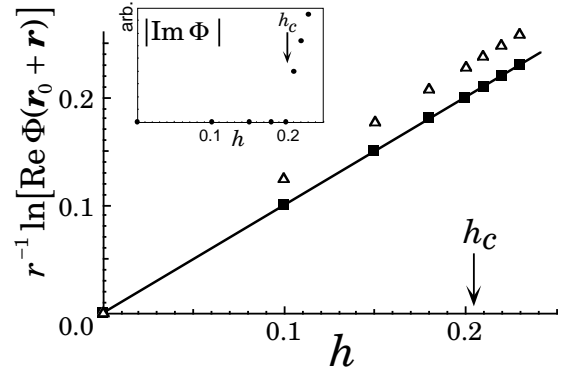,width=0.9\columnwidth}}
    \caption{The validity of the imaginary gauge transformation in the
      right wave function.  $r^{-1} \log |{\rm
        Re}\Phi(\bbox{r}_{0}+\bbox{r})|$ is plotted as a function of
      the imaginary vector potential $h$ where $\bbox{r}_{0}$ is the
      localization center.  The value is normalized by that at $h=0$.
      Squares refer to the values at $r=1$, and open triangles to
      those at $r=9$.  A solid line is what is expected by the
      imaginary gauge transformation.  Inset: $|{\rm Im}
      \Phi(\bbox{r}_{0}+\bbox{r})|$ for $r=1$ as a function
      of $h$. A jump at $h_{c}$ is clearly observed.  }
  \label{fig:WF}
\end{figure}

To justify the statement above, Fig.~\ref{fig:WF} illustrates the
spatial dependence of the right eigen wave function $\phi(\bbox{r})$
of the corresponding levels of Fig.~\ref{fig:DOS}.  When we look at
the amplitude next to a localization center, it evolves precisely in
the form predicted by the imaginary gauge transformation (the solid
line in the figure).  However, such dependence fails appreciably at a
point sufficiently far away from a localization center ($r=9$ on a
$20\times 20$ lattice).  The deviation is responsible for the level
shift observed in $h<h_{c}$, and has nothing to do with any precursor
to delocalization.  Simultaneously, Fig.~\ref{fig:WF} reveals clearly
that it is almost impossible, at least numerically, to detect the
point $h_{c}$ by examining the spatial dependence of the wavefunction
amplitude, in contrast to the one-dimensional system~\cite{Hatano98}.
The overall shape is barely changed through the critical value $h_{c}$
since the contribution of the imaginary part is too small to give a
visible deviation (see the inset).  It is remarked that on a few
occasions some abrupt movement of the localized center is observed
before reaching $h_{c}$ without any finite imaginary part.  We still
find, however, the spatial dependence following the gauge
transformation close to a localization center both before and after
it, so the phenomenon should be attributed as well to an artifact of
finite system size~\cite{Kuwae01b}.


To obtain $h_{c}$ from numerical data, we need to distinguish between
energies with vanishing and non-vanishing imaginary parts.  Such type
of estimate is usually hard in numerics, but in our problem, we
can unambiguously identify it, because complex
energy states always emerge by pairs (see Fig.\ref{fig:shiftedE}).

Following the procedure leading to Eq.~(\ref{eq:newrel}), we find
$h_{c}$ for a given configuration of the random potential, and take
the ensemble average to obtain the inverse localization length.  Our
numerical results are compared with the ones obtained from the
recursion method combined with finite scaling
analysis~\cite{MacKinnon83}, which is known among the most reliable
estimate of the localization length.  Since our numerical parameters
reside in a strongly localized region, we take the Lyapunov exponent
directly as the inverse localization length with little finite size
correction.  

\begin{figure}
  \centerline{\psfig{file=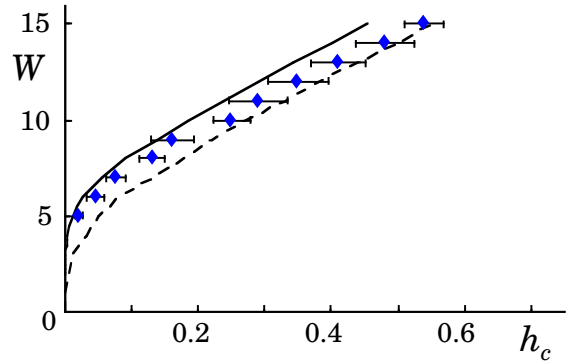,width=0.9\columnwidth}}
\caption{ $\langle h_{c}\rangle$ versus the disorder strength $W$ to
  confirm Eq.~(\protect\ref{eq:newrel}).  Data points ($\diamond$) are
  obtained from $40$ ensembles of a $20 \times 20$ lattice.  The solid
  line is the inverse localization length for a $\infty\times \infty$
  lattice, and the dashed line for a $16\times \infty$ one.
  Excerpted from Ref.~\protect\cite{MacKinnon83}.}
\label{fig:hc-xi}
\end{figure}

Figure~\ref{fig:hc-xi} shows the ensemble-averaged value $\langle
h_{c}\rangle $ and its variance from a $20\times 20$ lattices,
compared with the inverse localization length for a $\infty \times
\infty$ lattice denoted by the solid line (for $16\times\infty$
lattice by the dashed line) excerpted from Ref.~\cite{MacKinnon83}.
The agreement between $\langle h_{c} \rangle$ and the inverse
localization is clear, and the variance $\delta h_{c}$ is relatively
small, considering the fact we work on a comparatively small system
size and ensembles.  This shows the effectiveness of the present
approach.  Our data points show a systematic drift, though relatively
small, from the $\infty\times\infty$ line to the $16\times \infty$
line in approaching $h_{c}\to 1$.  The tendency is consistent because
finite size effect becomes increasingly prominent in this limit.  It
is emphasized that the agreement is substantial and the ability to
predict the localization length quantitatively is the virtue of
Eq.~(\ref{eq:newrel}), while other numerical objects such as the
participation ratio are often lacking.

As for the nature of the two-dimensional non-Hermitian delocalization
transition, our view based on the present numerical results is as
follows.  By attributing the level shift in the region $h<h_{c}$ to
finite size effect, it is natural to infer that the transition is not
continuous but abrupt in the limit of infinite size at least for a
given configuration.  By {\em assuming} further the self-averaging
property, one is tempted to reach the conclusion that the
two-dimensional non-Hermitian delocalization transition is abrupt for
bulk systems.  It should be recalled, however, that in our particular
choice of averaging procedure, we hardly respect the additional
symmetry present at the band center.  So we believe as a possible
scenario, that the transition away from the band center is
discontinuous but continuous at the band center, which is consistent
both with the nonperturbative analysis by the
supermatrix~\cite{Kolesnikov00} and with the symmetry argument
applicable to the band center~\cite{Mudry98}.  However, the present
results are not sufficient to conclude this clearly, and it requires
further studies in the future to draw a decisive conclusion on the
issue.


In conclusion, we have shown for the first time how the
two-dimensional localization can be probed by using the non-Hermitian
transition induced by the imaginary vector potential, proposing a
practical procedure to make an estimation of the inverse localization
length.  By checking the level shifting through the transition, it has
been understood clearly that the two-dimensional non-Hermitian
delocalization transition occurs through merging adjacent levels.  It
is true at present that the evaluation of the localization length by
using the non-Hermitian effect has not yet fulfilled such accuracy as
the recursion method enjoys, and it needs much more labor or fine
tuning to find $h_{c}$.  However, it surely serves as a useful
alternative that one may be able to apply even in experimental
settings.  Furthermore, as the present method need not increase the
system size, it may be advantageous in numerical studies on more
complicated systems such as the disordered interacting systems.


We acknowledge the support in the initial stage of the research
from the Grant-in-Aid for Scientific research No.11216204 by Japanese
Ministry of Education.

%


\begin{references}
  
\bibitem{Hatano9697} N. Hatano and D.~R. Nelson, Phys. Rev. Lett. {\bf
    77}, 570 (1996); Phys. Rev. B {\bf 56}, 8651 (1997).

\bibitem{Brouwer97} P.~W. Brouwer, P.~G. Silvestrov, and C.~W.~J.
  Beenakker, Phys. Rev. B {\bf 56}, R4333 (1997).
   
 \bibitem{Goldsheid98} I.~Y. Goldsheid and B.~A. Khruzhenko, Phys.
   Rev. Lett. {\bf 80}, 2897 (1998).

\bibitem{Silvestrov98} P.~G. Silvestrov, Phys. Rev. B {\bf 58}, R10111
  (1998).
  
\bibitem{Freilikher94} V. Freilikher, M. Pustilik and I. Yurkevich,
  Phys. Rev. Lett. {\bf 73}, 810 (1994); Phys. Rev. B{\bf 50}, 6017
  (1994).  

\bibitem{randomFP} D. Fisher {\it et al}., Phys. Rev. A {\bf 31}, 3841
  (1985); V. E. Kravtsov I. V. Lerner, V. I. Yudson, JETP {\bf 64},
  336 (1986).

\bibitem{Bouchaud95} J. P. Bouchaud, M. Mezard, and G. Parisi,
  Phys. Rev. E {\bf 52}, 3656 (1995).  

\bibitem{Mudry98b} C. Mudry {\it et al.}, Phys. Rev. B {\bf 58}, 13539
  (1998).

\bibitem{Mudry98} C. Mudry, B. D. Simons and A. Altland,
  Phys. Rev. Lett. {\bf 80}, 4257 (1998).

\bibitem{Kolesnikov00} A. V. Kolesnikov and K. B. Efetov, Phys. Rev.
  Lett. {\bf 84}, 5600 (2000).
  
\bibitem{note1} By convention of numerical studies, the localization
  length is referred to that along the lattice direction.  So the
  choice of $\bbox{h}=h(1,1)$ makes the $\sqrt{2}$ factor disappear in the
  following Eq.~(\ref{eq:newrel}).

\bibitem{GurarieBrezin98} V. Gurarie and A. Zee, cond-mat/9802042
  (unpublished); E. Br\'{e}zin and A. Zee, Nucl. Phys. {\bf B509}, 599
  (1998).  

\bibitem{Efetov97}
K.~B. Efetov, Phys.  Rev.  Lett.  {\bf 79}, 491 (1997).  
  

\bibitem{MacKinnon83} A. MacKinnon and B. Kramer, Z. Phys. {\bf B53},
  1 (1983).

  
\bibitem{Hatano98} N. Hatano and D.~R.  Nelson, Phys. Rev. B {\bf 58},
  8384 (1998).

\bibitem{Kuwae01b} T. Kuwae and N. Taniguchi (in preparation).

\end{references}


\widetext

\end{document}